\documentclass[aps,twocolumn,showpacs,nofootinbib,letterpaper]{revtex4}

\usepackage{graphicx}
\usepackage{dcolumn}
\usepackage{amsmath}
\usepackage{xspace}
\usepackage{bm}

\newcommand{\nuc}[2]{${}^{#1}$#2\xspace}
\renewcommand{\today}{\number\day\space\ifcase\month\or January\or 
 February\or March\or April\or May\or June\or July\or August\or 
 September\or October\or November\or December\fi\space\number\year}
\newcommand{\mco}[1]{\multicolumn{1}{c}{#1}}

\begin{document}

\setlength{\oddsidemargin}{0  pt} 

\title{Measurement of the Solar Neutrino Capture Rate \\
       by the Russian-American Gallium Solar Neutrino Experiment \\
       During One Half of the 22-Year Cycle of Solar Activity}

\author{J.\,N.\,Abdurashitov, V.\,N.\,Gavrin, S.\,V.\,Girin,
V.\,V.\,Gorbachev, P.\,P.\,Gurkina, T.\,V.\,Ibragimova, A.\,V.\,Kalikhov,
N.\,G.\,Khairnasov, T.\,V.\,Knodel, I.\,N.\,Mirmov, A.\,A.\,Shikhin,
E.\,P.\,Veretenkin, V.\,M.\,Vermul, V.\,E.\,Yants, and G.\,T.\,Zatsepin}
\affiliation{Institute for Nuclear Research, Russian Academy of Sciences,
117312 Moscow, Russia}

\author{T.\,J.\,Bowles and W.\,A.\,Teasdale}
\affiliation{Los Alamos National Laboratory, Los Alamos, New Mexico
87545, USA}

\author{J.\,S.\,Nico}
\affiliation{National Institute of Standards and Technology, Stop 8461,
Gaithersburg, Maryland 20899, USA}

\author{B.\,T.\,Cleveland, S.\,R.\,Elliott, and J.\,F.\,Wilkerson}
\affiliation{University of Washington, Seattle, Washington 98195, USA}

\author{(The SAGE Collaboration)}
\noaffiliation

\date{\today}

\begin{abstract}

We present the results of measurements of the solar neutrino capture
rate in gallium metal by the Russian-American Gallium Experiment SAGE
during slightly more than half of a 22-year cycle of solar activity.
Combined analysis of the data of 92~runs during the 12-year period
January 1990 through December 2001 gives a capture rate of solar
neutrinos with energy more than 233~keV of $70.8 ^{+5.3} _{-5.2}$
(stat.)  $^{+3.7} _{-3.2}$ (syst.)~SNU.  This represents only slightly
more than half of the predicted standard solar model rate of 128~SNU.
We give the results of new runs beginning in April 1998 and the
results of combined analysis of all runs since 1990 during yearly,
monthly, and bimonthly periods.  Using a simple analysis of the SAGE
results combined with those from all other solar neutrino experiments,
we estimate the electron neutrino $pp$ flux that reaches the Earth to
be $(4.6 \pm 1.1) \times 10^{10}$/(cm$^2$-s).  Assuming that neutrinos
oscillate to active flavors the $pp$ neutrino flux emitted in the
solar fusion reaction is approximately $(7.7 \pm 1.8) \times
10^{10}$/(cm$^2$-s), in agreement with the standard solar model
calculation of $(5.95 \pm 0.06)\times 10^{10}$/(cm$^2$-s).

\end{abstract}


\maketitle

\markboth {\hfill SAGE JETP \hfill \today}
          {\today \hfill SAGE JETP \hfill}

\section{Introduction}

There has been outstanding progress in solar neutrino research in the
last several years.  The large water Cherenkov detectors
SuperKamiokande (SK)~\cite{KAM01} and the Sudbury Neutrino Observatory
(SNO)~\cite{MAC00} have begun to measure the high-energy solar
neutrinos from \nuc{8}{B} decay in real time with a high counting
rate.  The results from these two neutrino telescopes of a new
generation add significant information to the existing results from
the chlorine \cite{CLE98} and gallium \cite{ABD99,HAM99} radiochemical
experiments and the Kamiokande experiment \cite{FUK96}.  The elastic
scattering results of SK combined with the charged current results
from SNO indicate that, in addition to electron neutrinos, active
neutrinos of other flavors are reaching the Earth from the Sun.  The
combined analysis of results of all these experiments gives convincing
evidence that part of the electron neutrinos formed in thermonuclear
reactions in the Sun change their flavor on their way to the Earth.

Further measurement of the details of flavor conversion for solar
neutrinos requires a new generation of neutrino telescopes that will
be sensitive to the spectrum below 2~MeV, a region that contains the
proton-proton $(pp)$ and CNO continua as well as the \nuc{7}{Be} and
$pep$ lines.  Although there are many promising ideas for real-time
low-energy neutrino detectors \cite{LONU}, at the present time only
the radiochemical gallium experiments are able to measure and monitor
this part of the solar neutrino spectrum.  The 233-keV threshold of
the reaction $^{71}$Ga$(\nu_e,e^-)^{71}$Ge \cite{KUZ65} enables one to
measure the $pp$ neutrinos, the principal component of the solar
neutrino spectrum.  If we exclude exotic hypotheses, the rate of the
$pp$ reaction is directly related to the solar luminosity and is
insensitive to alterations in the solar models that influence the
subsequent reactions in the solar fusion chain.

The neutrino capture rate in \nuc{71}{Ga} predicted by the Standard
Solar Model (SSM) \cite{BAH00} is $128^{+9}_{-7}$~SNU, with the main
contribution of 69.7~SNU from the $pp$ neutrinos~\footnote{1~SNU =
1~interaction/s in a target that contains 10$^{36}$ atoms of the
neutrino absorbing isotope.}.  Contributions by \nuc{7}{Be} and
\nuc{8}{B} neutrinos are 34.2~SNU and 12.1~SNU, respectively.  The
insensitivity of Ga to variation in the solar model is seen in the
independently calculated result of 127.2~SNU~\cite{TUR98} for the
total capture rate.

From the results of the SNO and SK experiments we now have a high
accuracy measurement of the \nuc{8}{B} flux that reaches the Earth and
its electron neutrino component.  In the near future it is expected
that the KamLAND experiment~\cite{Kamland} will greatly restrict the
range of possible electron neutrino mixing parameters.  This data,
combined with results from the Borexino experiment~\cite{Borexino},
will give us a good measurement of the \nuc{7}{Be} flux from the Sun.
By subtracting the \nuc{7}{Be} and \nuc{8}{B} components from the
total signal in the Ga experiment, we will obtain a measurement of a
fundamental astrophysical parameter --- the neutrino flux from the
proton-proton fusion reaction.  There will be a slight contamination
from the CNO and $pep$ neutrinos but this can be removed with the
information from the Cl experiment.  A rough estimate of the $pp$
neutrino flux using the information that we have now is presented in
the next to last section.  Since in the immediate future only the Ga
experiments provide this measurement, it is very important that both
SAGE \cite{ABD99} and GALLEX's successor GNO \cite{GNO00} continue to
operate so as to improve the accuracy of their results.

The experimental layout and procedures, including extraction of
germanium from gallium, counting of \nuc{71}{Ge}, and data analysis,
are described in detail in our article ``Measurement of the solar
neutrino capture rate with gallium metal'' in Physical Review C
\cite{ABD99}.  That article gives the SAGE results for the period
January 1990 through December 1997.  Since it contains a complete
description of the experiment and the analysis techniques, we refer
the reader who wishes further detail to that publication.  Here we
briefly discuss the main principles of the experiment, give the
statistical analysis of the data from 1998--2001, present new results
for some systematic uncertainties, and conclude with the current
implications of the SAGE result for solar and neutrino physics.

\section{Overview of the SAGE Experiment}
\subsection{The Laboratory of the Gallium Germanium Neutrino Telescope}

The SAGE experiment is situated in a specially built deep underground
laboratory \cite{GGNT} at the Baksan Neutrino Observatory (BNO) of the
Institute for Nuclear Research of the Russian Academy of Sciences in
the northern Caucasus mountains.  It is located 3.5~km from the
entrance of a horizontal adit excavated into the side of Mount
Andyrchi.  The main chamber of the laboratory is 60~m long, 10~m wide,
and 12~m high.  The rock gives an overhead shielding equivalent to
4700~m of water and reduces the muon flux by a factor of $10^7$.  The
measured muon flux is $(3.03 \pm 0.10) \times 10^{-9}$/(cm$^2$-s)
\cite{muonflux}.  To reduce neutron and gamma backgrounds from the
rock, the laboratory is entirely lined with 60~cm of low-radioactivity
concrete with an outer 6~mm steel shell.  The flux of neutrons with
energy (1--11) MeV is less than $2.3 \times 10^{-7}$/(cm$^2$-s)
\cite{neutronflux}.  All facilities required by the experiment are in
this underground area, with additional rooms devoted to chemistry,
counting, and a low-background solid-state Ge detector.  Subsidiary
measurements are made in a general laboratory building outside the
adit.

\subsection{Experimental Procedures}

The gallium target used at the present time for measurements is about
50~t.  It is in the form of liquid metal and is contained in 7
chemical reactors.  A measurement of the solar neutrino capture rate,
which we call a ``run'', begins by adding to the gallium a stable Ge
carrier.  The carrier is a Ga-Ge alloy with a known Ge content of
approximately 350~$\mu$g and is distributed equally among all
reactors.  The reactor contents are stirred thoroughly to disperse the
Ge throughout the Ga mass.  After a typical exposure interval of
4~weeks, the Ge carrier and \nuc{71}{Ge} atoms produced by solar
neutrinos and background sources are chemically extracted from the Ga
using procedures described elsewhere \cite{ABD99,ABD94}.  The final
step of the chemical procedure is the synthesis of germane (GeH$_4$),
which is used as the proportional counter fill gas with an
admixture of (80--90)\%~Xe.  The total efficiency of extraction is the
ratio of mass of Ge in the germane to the mass of initial Ge carrier
and is typically in the range of (80--90)\%.  The systematic
uncertainty in this efficiency is 3.4\%, mainly arising from
uncertainties in the mass of added and extracted carrier.  Each
extraction dissolves $\sim$0.1\% of the gallium.  Ga metal is
regenerated from the accumulated extracted solutions, purified, and
subsequently will be returned to the target.

The proportional counter is placed in the well of a NaI detector that
is within a large passive shield and is counted for a typical period
of 4--6~months.  \nuc{71}{Ge} decays via electron capture to
\nuc{71}{Ga} with a half-life of 11.43~d.  Low energy $K$- and
$L$-shell Auger electrons and x~rays from electron shell relaxation
produce a nearly point-like ionization in the counter gas, resulting
in a fast rise time for the pulse from the counter.  In contrast, most
background events leave an extended trail of ionization, producing a
slower rise time.  A candidate \nuc{71}{Ge} event must thus have an
energy within a window around an $L$ or $K$ peak and must also have a
rise time consistent with point-like ionization.  In addition, since
\nuc{71}{Ge} decays without gamma emission, this event must not be
coincident with an event in the NaI detector.

\begin{table*} \caption{Parameters for all runs beginning with April
1998 that are used to give the solar neutrino result.  The efficiency
values include the reduction for the energy cut (0.9815), for the rise
time cut (0.95), and for the faulty data acquisition program for runs
prior to 2000 (0.76).  The `Peak ratio' is the factor by which an
\nuc{55}{Fe} calibration must be multiplied to correct the energy
scale; it is 1.0 for a counter with no polymerization.}
\label{runparameters}
\begin{tabular*}{\hsize}{l @{\extracolsep{\fill}} c c c c c c d c c c c}
\hline
\hline
            & Mean       & Exposure  & Ga \\
Exposure   & exposure   & time      & mass   & Extraction & Counter &
Pressure  & \multicolumn{1}{c}{Percent} & Operating & $K$-peak   & $L$-peak
& Peak   \\
  date      & date       & (days)    & (t) & efficiency & name    & (mm Hg)  &
\multicolumn{1}{c}{GeH$_4$} & voltage   & efficiency & efficiency & ratio  \\
\hline
Apr. 98 & 1998.225 &  44.9 & 48.05 & 0.85 & A13    & 695 & 37.0 & 1480 & 0.243 & 0.219 & 1.01  \\
May  98 & 1998.347 &  30.0 & 51.17 & 0.91 & LY4    & 690 & 29.5 & 1366 & 0.238 & 0.245 & 1.00  \\
July 98 & 1998.477 &  45.6 & 51.06 & 0.90 & A12    & 680 & 32.0 & 1414 & 0.235 & 0.237 & 1.00  \\
Aug. 98 & 1998.611 &  45.7 & 50.93 & 0.89 & LA51   & 660 & 27.0 & 1356 & 0.234 & 0.244 & 1.04  \\
Oct. 98 & 1998.745 &  45.8 & 50.81 & 0.92 & A13    & 680 & 32.0 & 1404 & 0.244 & 0.212 & 1.00  \\
Nov. 98 & 1998.883 &  45.8 & 50.68 & 0.92 & LY4    & 680 & 26.5 & 1322 & 0.238 & 0.244 & 1.00  \\
Jan. 99 & 1999.014 &  44.7 & 50.54 & 0.92 & A12    & 700 & 30.0 & 1398 & 0.239 & 0.241 & 1.00  \\
Feb. 99 & 1999.130 &  38.7 & 50.43 & 0.89 & LA51   & 705 & 11.0 & 1194 & 0.248 & 0.234 & 1.05  \\
Apr. 99 & 1999.279 &  51.7 & 50.29 & 0.89 & A13    & 665 & 13.5 & 1206 & 0.253 & 0.231 & 1.05  \\
June 99 & 1999.417 &  46.7 & 50.17 & 0.87 & LY4    & 670 & 11.0 & 1140 & 0.246 & 0.239 & 1.00  \\
July 99 & 1999.551 &  45.7 & 50.06 & 0.90 & LA116  & 635 & 12.5 & 1164 & 0.243 & 0.244 & 1.03  \\
Sep. 99 & 1999.685 &  45.7 & 49.91 & 0.91 & LA51   & 660 & 11.5 & 1172 & 0.242 & 0.238 & 1.05  \\
Oct. 99 & 1999.801 &  38.7 & 49.78 & 0.90 & A13    & 665 & 12.5 & 1186 & 0.254 & 0.202 & 1.01  \\
Jan. 00 & 2000.035 &  28.8 & 49.59 & 0.91 & LA51   & 700 & 13.5 & 1224 & 0.324 & 0.310 & 1.05  \\
Feb. 00 & 2000.127 &  30.7 & 49.48 & 0.83 & LY4    & 646 & 10.4 & 1130 & 0.320 & 0.316 & 1.01  \\
Mar. 00 & 2000.207 &  28.8 & 49.42 & 0.91 & A13    & 665 & 14.5 & 1206 & 0.332 & 0.329 & 1.10  \\
May  00 & 2000.359 &  30.7 & 49.24 & 0.92 & LA116  & 705 & 14.0 & 1244 & 0.329 & 0.315 & 1.03  \\
June 00 & 2000.451 &  33.7 & 49.18 & 0.84 & LA51   & 652 & 12.0 & 1160 & 0.317 & 0.314 & 1.03  \\
July 00 & 2000.540 &  32.0 & 49.12 & 0.92 & LY4    & 670 & 13.8 & 1182 & 0.321 & 0.316 & 1.01  \\
Aug. 00 & 2000.626 &  31.3 & 49.06 & 0.73 & A13    & 707 &  9.5 & 1176 & 0.343 & 0.321 & 1.08  \\
Sep. 00 & 2000.704 &  27.7 & 49.00 & 0.89 & A12    & 690 & 14.7 & 1224 & 0.324 & 0.312 & 1.00  \\
Oct. 00 & 2000.796 &  30.7 & 48.90 & 0.84 & LA116  & 734 &  9.4 & 1188 & 0.337 & 0.303 & 1.03  \\
Nov. 00 & 2000.876 &  28.7 & 48.84 & 0.93 & LA51   & 680 & 11.9 & 1196 & 0.345 & 0.330 & 1.03  \\
Dec. 00 & 2000.958 &  30.7 & 48.78 & 0.93 & LY4    & 697 & 12.0 & 1174 & 0.327 & 0.312 & 1.02  \\
Feb. 01 & 2001.122 &  29.8 & 41.11 & 0.87 & LA116  & 687 &  9.2 & 1144 & 0.330 & 0.314 & 1.04  \\
Mar. 01 & 2001.214 &  33.4 & 48.53 & 0.92 & LA51   & 635 & 13.5 & 1180 & 0.314 & 0.317 & 1.02  \\
Apr. 01 & 2001.290 &  22.7 & 48.43 & 0.90 & YCT1   & 695 & 13.1 & 1210 & 0.344 & 0.333 & 1.00  \\
May  01 & 2001.373 &  31.7 & 48.37 & 0.88 & YCT2   & 625 & 14.9 & 1178 & 0.332 & 0.342 & 1.00  \\
June 01 & 2001.469 &  31.7 & 48.27 & 0.92 & YCT3   & 678 & 12.2 & 1190 & 0.342 & 0.334 & 1.00  \\
July 01 & 2001.547 &  23.7 & 48.17 & 0.93 & LA116  & 690 & 12.7 & 1196 & 0.328 & 0.315 & 1.03  \\
Aug. 01 & 2001.624 &  28.7 & 48.11 & 0.59 & A12    & 768 &  7.2 & 1148 & 0.340 & 0.302 & 1.00  \\
Sep. 01 & 2001.701 &  27.7 & 48.06 & 0.90 & YCT1   & 665 & 15.0 & 1204 & 0.338 & 0.337 & 1.00  \\
Oct. 01 & 2001.793 &  30.7 & 47.96 & 0.88 & YCT2   & 758 & 12.2 & 1210 & 0.354 & 0.326 & 1.00  \\
Nov. 01 & 2001.887 &  34.8 & 47.91 & 0.92 & YCT3   & 685 & 14.2 & 1210 & 0.342 & 0.335 & 1.00  \\
Dec. 01 & 2001.955 &  22.8 & 47.86 & 0.86 & YCT4   & 685 & 11.4 & 1176 & 0.344 & 0.333 & 1.00  \\
\hline
\hline
\end{tabular*}
\end{table*}

The data acquisition electronics have evolved over the course of SAGE.
During the first two years a hardware measure of the pulse rise time
was used --- the amplitude of the differentiated pulse (ADP) technique.
This method suffices well in the $K$ peak (10.4~keV) but was found to
be inadequate for the $L$ peak (1.2~keV), which is more sensitive to
electronic drifts and has higher background.  In 1992 an 8-channel
counting system with a 1-GHz digital oscilloscope was implemented,
which permits off-line analysis of the event waveforms.  The digitized
pulse is fit to a functional form \cite{ELL90} which gives the energy
deposited during the event and the time duration $T_N$ over which the
ionization arrives at the anode wire.  All $L$-peak results and the
vast majority of $K$-peak results are obtained from this waveform
analysis.

After filling, counters are calibrated with an \nuc{55}{Fe} source
(5.9~keV) through a window in the Fe cathode.  Typically, they are
again calibrated after 3~days of operation and approximately every two
weeks subsequently.  Calibrations are also made with a \nuc{109}{Cd}
source whose gamma rays penetrate the counter wall and fluoresce the
length of the Fe cathode, thus giving the $K$ x-ray peak from Fe at
6.4~keV.  This allows correction of the \nuc{71}{Ge} peak position due
to the accumulation of polymer deposits on the anode wire which may
occur after prolonged operation.  A \nuc{109}{Cd}+Se source, which
gives peaks at 1.4~keV and 11.2~keV, is also occasionally used to
check linearity.

The measure of energy is the integral of the pulse waveform for 800~ns
after pulse onset.  The \nuc{71}{Ge} peak position is based on the
\nuc{55}{Fe} calibration adjusted for polymerization and the energy
window (two full widths at half maximum) is set by the \nuc{55}{Fe}
resolution.  If the calibration centroid shifts between two
calibrations, the window for energy selection is linearly shifted in
time between the two calibrations.  Typical gain shifts are of the
order of a few percent which leads to a calculated uncertainty in the
counting efficiency of no more than -3.1\%.

\begin{figure*}
\begin{center}
\includegraphics[width=0.8\textwidth]{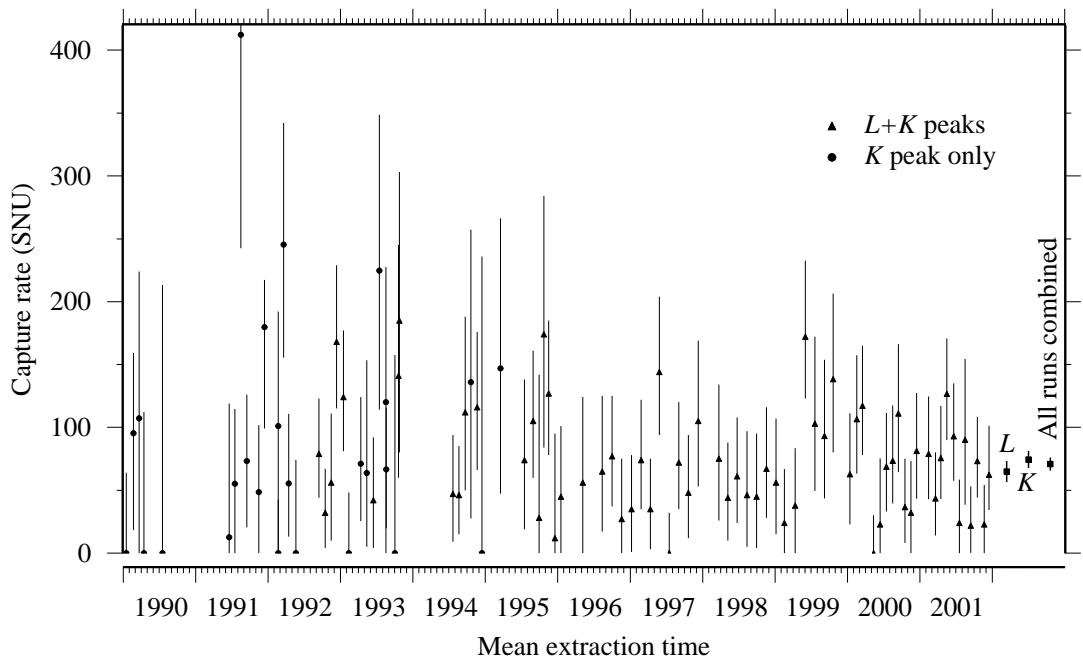}
\end{center}
\caption{Capture rate for all SAGE extractions as a function of time.  
Error bars are statistical with 68\% confidence.  The combined result
of all runs in the $L$ peak, the $K$ peak, and both $L$ and $K$ peaks
is shown on the right side.  The last 3 runs are still counting and 
their results are preliminary.}
\label{ratevstime}
\end{figure*}

To determine the rise time windows for \nuc{71}{Ge} events,
proportional counters filled with $^{71}$GeH$_4$ are measured in
each of the counting channels.  All events inside the energy windows
of the $K$ and $L$ peaks are selected, and the rise time $T_N$ of each
event is calculated \cite{ELL90}.  The rise time values are arranged
in ascending order and limits are set such that 5\% of the events are
excluded.  For a counter filled with 30\% GeH$_4$ this leads to $K$-
and $L$-peak event selection limits on $T_N$ of 0.0~ns to 18.4~ns and
0.0~ns to 10.0~ns, respectively.  The upper limits are reduced to
13.2~ns and 9~ns for the $K$ and $L$ peaks, respectively, when the
GeH$_4$ concentration is in the range of 7\% to 15\%.  The variation
with counter filling and electronics channel results in an uncertainty
in the efficiency of about $\pm$1\%.  The small loss in counting
efficiency associated with this cut leads to a significant reduction
in the number of background events.

Table~\ref{runparameters} gives the parameters of the 35 runs from
1998--2001 that are used for solar neutrino measurement.

\section{Statistical Analysis of Solar Data}

The selection criteria result in a group of events from each
extraction that are candidate \nuc{71}{Ge} decays.  These events are
fit to a maximum likelihood function \cite{CLE83}, assuming that they
originate from an unknown but constant-rate background and the
exponentially decaying rate of \nuc{71}{Ge}.

Two time cuts are made on the counting intervals to reduce the effect
of radon and radon daughters, which can give pulses that mimic
\nuc{71}{Ge}.  To minimize the effect of Rn that enters the interior
of the passive shield when it is opened for counter calibration, we
delete the first 2.6~h of counting time in the $K$ and $L$ peaks after
shield openings.  Another very dangerous background occurs if even a
few atoms of \nuc{222}{Rn} enter the counter during filling.  Most
decays of Rn inside the counter give slow pulses at a saturated energy
beyond the \nuc{71}{Ge} peaks, but approximately 8\% of the pulses
from Rn and its daughters make fast pulses that are indistinguishable
from those of \nuc{71}{Ge}.  Since the radon decay chain takes on
average 50~min to reach the long-lived isotope \nuc{210}{Pb}, deleting
15~min before and 3~h after each saturated pulse removes most of these
internal Rn events.
\begin{table*}
\caption{Results of combined analysis of $K$-peak and $L$-peak events
for all runs after 1997.  See {\protect \cite{CLE982}} for the
definition and interpretation of $Nw^2$.  The accuracy of the Monte
Carlo-determined goodness of fit probability is $\sim$1.5\% for each
individual run and $\sim$4\% for the combination of all runs.  The 
last 3 runs are still counting and their results are preliminary.}
\label{recentruns}
\begin{tabular*}{0.75\hsize}{l @{\extracolsep{\fill}} d d d r
      @{\extracolsep{0pt}--} d @{\extracolsep{\fill}} c d}
\hline
\hline
          & \mco{Number of} & \mco{Number   }  \\
Exposure  & \mco{candidate} & \mco{fit to   }  & \mco{Best fit}
   & \multicolumn{2}{c}{68\% conf.}  &               & \mco{Prob.} \\
  date    & \mco{events   } & \mco{$^{71}$Ge}  & \mco{  (SNU) }
   & \multicolumn{2}{c}{range (SNU)} & \mco{$Nw^2$}  & \mco{(\%)}  \\
\hline
 Apr. 98  &    39 &    5.4 &   75 &   26 &  134 & 0.052 &   72 \\
 May  98  &    23 &    3.4 &   44 &   10 &   88 & 0.051 &   68 \\
 July 98  &    22 &    4.8 &   61 &   24 &  108 & 0.065 &   52 \\
 Aug. 98  &    33 &    3.6 &   46 &    5 &   97 & 0.039 &   84 \\
 Oct. 98  &    40 &    3.8 &   45 &    4 &   95 & 0.028 &   95 \\
 Nov. 98  &    32 &    5.9 &   67 &   28 &  116 & 0.101 &   30 \\
 Jan. 99  &    21 &    4.5 &   56 &   15 &  107 & 0.036 &   84 \\
 Feb. 99  &    16 &    1.6 &   24 &    0 &   67 & 0.114 &   28 \\
 Apr. 99  &    10 &    1.8 &   38 &    5 &   83 & 0.105 &   36 \\
 June 99  &    14 &   12.9 &  172 &  123 &  232 & 0.048 &   80 \\
 July 99  &    17 &    5.5 &  103 &   49 &  172 & 0.118 &   20 \\
 Sep. 99  &    20 &    7.1 &   93 &   43 &  154 & 0.099 &   28 \\
 Oct. 99  &    16 &   10.0 &  138 &   80 &  206 & 0.066 &   56 \\
 Jan. 00  &    24 &    5.4 &   63 &   23 &  111 & 0.060 &   59 \\
 Feb. 00  &    21 &    9.1 &  107 &   63 &  157 & 0.058 &   55 \\
 Mar. 00  &    19 &   10.1 &  117 &   78 &  165 & 0.046 &   79 \\
 May  00  &    15 &    0.0 &    0 &    0 &   32 & 0.143 &   40 \\
 June 00  &    17 &    1.4 &   23 &    0 &   75 & 0.179 &   17 \\
 July 00  &    29 &    6.4 &   69 &   33 &  111 & 0.088 &   34 \\
 Aug. 00  &    14 &    5.2 &   74 &   39 &  117 & 0.086 &   33 \\
 Sep. 00  &    30 &    9.2 &  111 &   64 &  166 & 0.093 &   24 \\
 Oct. 00  &    14 &    3.0 &   37 &    8 &   75 & 0.020 &   99 \\
 Nov. 00  &    25 &    2.9 &   32 &    0 &   73 & 0.208 &    9 \\
 Dec. 00  &    27 &    7.6 &   81 &   43 &  127 & 0.062 &   68 \\
 Feb. 01  &    21 &    6.3 &   79 &   43 &  125 & 0.088 &   34 \\
 Mar. 01  &    18 &    3.8 &   44 &   14 &   80 & 0.120 &   24 \\
 Apr. 01  &    17 &    6.7 &   76 &   43 &  117 & 0.074 &   45 \\
 May  01  &    21 &   11.9 &  127 &   90 &  171 & 0.088 &   31 \\
 June 01  &    20 &    9.4 &   93 &   57 &  135 & 0.025 &   96 \\
 July 01  &     9 &    2.0 &   24 &    0 &   58 & 0.033 &   92 \\
 Aug. 01  &    21 &    5.4 &   90 &   38 &  155 & 0.065 &   57 \\
 Sep. 01  &    10 &    2.1 &   22 &    0 &   53 & 0.139 &   18 \\
 Oct. 01  &    12 &    7.5 &   73 &   44 &  109 & 0.082 &   41 \\
 Nov. 01  &    15 &    2.6 &   23 &    0 &   54 & 0.084 &   38 \\
 Dec. 01  &     9 &    5.2 &   62 &   34 &  101 & 0.063 &   70 \\
\hline
 Combined &   711 &  191.8 &   67 &   60 &   74 & 0.080 &   42 \\
\hline
\hline
\end{tabular*}
\end{table*}

For each individual extraction, the best estimate of the \nuc{71}{Ge}
production rate is found by maximizing the likelihood function.  The
statistical uncertainty in the production rate is found by integrating
the likelihood function over the background rate to give a function of
production rate only, and then finding the minimum range in this rate
that includes 68\% of the area under the curve.  This procedure is done
separately for events in the $L$ and $K$ peaks.  The small variation
in the Earth-Sun distance (no greater than 3\%) due to the
eccentricity of the Earth's orbit is taken into account.  The best
estimate for a combination of runs (and also for the $L+K$ combination
of a single run) is obtained by multiplying the individual likelihood
functions, requiring that the production rate per unit mass of Ga be
equivalent but allowing the background rate to be different for each
component of the function.  The global maximum of the product function
is then found and the 68\% confidence region for the production rate
is set by finding where the function has decreased from its value at
the maximum by the factor 0.606, all other variables being maximized.
The results of the combined analysis of recent extractions are given
in Table~\ref{recentruns} and the results of all runs of SAGE are
plotted in Fig.~\ref{ratevstime}.

After publication of Ref.~\cite{ABD99}, which reported measurements
from January 1990 through December 1997, it was found that a faulty
data acquisition program was used in the period from June 1996 through
December 1999.  At the beginning of this period it was necessary to
change the hardware that recognized coincidences between the NaI and
proportional counters.  The new hardware required a modification of
the acquisition program which introduced an error in the trigger
logic.  As a result $23.9 \pm 0.4 \text{ (stat.)}  \pm 0.5 \text{
(syst.)}$\% of triggers was lost.  This error artificially reduced the
results of individual runs counting during this period and slightly
influenced the overall result.  Corrected results were given in
\cite{GAV00}.

\section{Systematic Effects}

Table~\ref{systematics} summarizes the systematic effects that may
affect the measured solar neutrino capture rate.  These uncertainties
fall into three main categories: those associated with extraction
efficiency, with counting efficiency, and with backgrounds.  Some of
these effects were mentioned above and the others will be briefly
discussed here.

The counting efficiency and its uncertainty was determined by a series
of measurements with gas fillings of \nuc{71}{Ge}, \nuc{37}{Ar}, and
\nuc{69}{Ge}.  The uncertainty contains terms due to volume
efficiency, end effects, and gas efficiency.  Adding each of these
effects in quadrature gives a $\pm1.8\%$ uncertainty due to the
counters.

There exist small contributions to the \nuc{71}{Ge} signal by means
other than solar neutrinos.  Limits on the creation of \nuc{71}{Ge}
through the $(n,p)$ reaction on \nuc{71}{Ga} and by cosmic-ray muons
were obtained by measurement in the Ga chamber of both the fast
neutron \cite{neutronflux,Gavrin91b} and muon fluxes \cite{muonflux}.
Limiting values on the U and Th concentration in the Ga, which can
also make \nuc{71}{Ge}, have been determined by low background
counting in a Ge detector \cite{GAV86} and by glow discharge mass
spectrometry \cite{Evans}.  The inferred \nuc{71}{Ge} production rate
from the combination of all of these processes is no more than 1~SNU.

Rn decay in the vicinity of the proportional counters or inside the
counters themselves can also give events that mimic \nuc{71}{Ge}
decay.  To reduce such events Rn is purged from the volume near the
counters by a constant flow of gas from boiling liquid nitrogen and
special anti-Rn procedures are applied to purify the gas mixture at
the time of counter filling.  The influence of residual Rn was
quantified by studies in which Rn was added inside the counter
\cite{intradon} and the response of the counter to external gamma-rays
was measured \cite{extgamma}.  The systematic uncertainties on the
\nuc{71}{Ge} production rate for Rn that remains after the time cuts
were determined to be $<$0.2~SNU for internal Rn and 0.03~SNU for
external Rn.

\begin{table}
\caption{Summary of systematic effects and their uncertainties in SNU.
The values for extraction and counting efficiencies are based on a
rate of 70.8~SNU.}
\label{systematics}
\begin{tabular*}{\hsize}{l @{\extracolsep{\fill}} l c}
\hline
\hline
Extraction efficiency & Ge carrier mass         & $\pm$1.5   \\
                      & Extracted Ge mass       & $\pm$1.8   \\
                      & Residual carrier Ge     & $\pm$0.6   \\
                      & Ga mass                 & $\pm$0.2   \\
\\
Counting efficiency   & Counter effects         & $\pm$1.3   \\
                      & Gain shifts             & +2.2       \\
                      & Resolution              & -0.4,+0.5  \\
                      & Rise time limits        & $\pm$0.7   \\
                      & Lead and exposure times & $\pm$0.6   \\
\\
Backgrounds           & Neutrons                & $<$-0.02   \\
                      & U and Th                & $<$-0.7    \\
                      & muons                   & $<$-0.7    \\
                      & Internal radon          & $<$-0.2    \\
                      & External radon          &  0.0       \\
                      & Other Ge isotopes       & $<$-0.7    \\
\\
Total                 &                         & -3.2,+3.7  \\
\hline
\hline
\end{tabular*}
\end{table}

\nuc{68}{Ge} and \nuc{69}{Ge} can be produced by background processes
and their decays can be misinterpreted as \nuc{71}{Ge}.  \nuc{68}{Ge}
is of special concern as it decays solely by electron capture giving
events that are identical to those from \nuc{71}{Ge}.  It is mainly
produced by cosmic-ray muons and we can estimate the rate by scaling
the predicted \nuc{71}{Ge} rate of $0.012 \pm 0.06$ atoms/day in 60~t
of Ga~\cite{ABD99,Gavrin87} by the measured cross section ratio for
\nuc{68}{Ge} to \nuc{71}{Ge} production with 280-GeV muons of $2.2 \pm
0.1$~\cite{Cribier97}.  This gives $0.022 \pm 0.013$ atoms of
\nuc{68}{Ge} produced by muons per day in 50~t of Ga.  Since the
\nuc{68}{Ge} half life is 271~d, considerably longer than the usual
total counting time, these pulses will be distributed almost uniformly
during the counting period, but there will be a small excess during
the early part of counting.  Using typical values for exposure time of
30~d, Ga mass of 50~t, extraction efficiency of 0.9, counting
efficiency of 0.6 in the sum of $L$ and $K$ peaks, 150~d of counting,
and our average $L+K$ background rate of 0.175/day, simulations show
that a true \nuc{68}{Ge} production rate of 0.022/day leads to a false
\nuc{71}{Ge} production rate of 0.0085/day, which is equivalent to
0.05~SNU.

The isotope \nuc{69}{Ge} can be produced from Ga by several processes,
including the interaction of cosmic rays, \nuc{8}{B} solar neutrinos,
and protons on \nuc{71}{Ga}, where the protons are secondaries either
from $\alpha$~particles from internal radioactivity or from neutrons
from the surrounding rock.  The total \nuc{69}{Ge} production rate in
60~t of Ga is estimated to be 0.21~atoms/day~\cite{ABD99} with an
uncertainty of approximately 50\%.  Because most \nuc{69}{Ge} decays
are accompanied by gamma rays which are rejected by the NaI veto with
an efficiency of 90\%, and counting usually begins $\sim$1.5~d after
extraction, a time comparable with the \nuc{69}{Ge} half life of
1.6~d, we calculate that only 0.045~\nuc{69}{Ge} decays will be
observed in a typical run~\cite{ABD99}, a factor of 100 less than the
average number of detected \nuc{71}{Ge} decays.  The background effect
of \nuc{69}{Ge} is thus no more than 0.7~SNU.

These estimates of the counting rate of \nuc{68}{Ge} and \nuc{69}{Ge}
can be checked by searching for these isotopes in the large set of
\nuc{71}{Ge} counting data that we now have from solar runs.  The
procedures and efficiencies for selecting candidate events are
described in Ref.~\cite{Gorbachev}.  The \nuc{68}{Ge} production rate
determined in this way is $0.18^{+0.13}_{-0.12}$~atoms/day in 60~t of
Ga.  The central value is a factor of~7 more than the estimate given
above, but within the large error range the rates determined by the
two methods are in agreement.  Since the result from the cosmic-ray
production rate of \nuc{71}{Ge} and the cross section ratio has the
smaller uncertainty, we use it for our estimate of the \nuc{68}{Ge}
background.  There is some concern though that the predicted
cosmic-ray production rate of \nuc{71}{Ge} may be underestimated; a
more direct measurement would be desirable and this is considered
briefly in \cite{Gorbachev}.  A similar search for \nuc{69}{Ge} events
shows that the production rate in 60~t of gallium is less than
0.49~atoms/day, in agreement with the value given above.

\section{Results}

The global best fit capture rate for all data from January 1990 to
December 2001 (92 runs and 158 separate counting sets) is $70.8
^{+5.3} _{-5.2}$~SNU, where the uncertainty is statistical only.  In
the windows that define the $L$ and $K$ peaks there are 1723 counts
with 406.4 assigned by time analysis to \nuc{71}{Ge} (the total
counting live time is 29.5 years).  If one considers the $L$-peak and
$K$-peak data independently, the results are $64.8^{+8.5}_{-8.2}$~SNU
and $74.4^{+6.8}_{-6.6}$~SNU, respectively.  The agreement between the
two peaks serves as a strong check on the robustness of the event
selection criteria.  The total systematic uncertainty is determined by
adding in quadrature all the contributions given in
Table~\ref{systematics}.  Our overall result is thus $70.8 ^{+5.3
+3.7}_{-5.2 -3.2}$~SNU.  For comparison, the latest result of the GNO
experiment (including GALLEX) is $74.1 ^{+5.4 +4.0} _{-5.4
-4.2}$~SNU~\cite{GNO00}.  If we combine the SAGE statistical and
systematic uncertainties in quadrature, the result is $70.8
^{+6.5}_{-6.1}$~SNU.

\subsection{Tests of \nuc{71}{Ge} Extraction Efficiency}

The validity of this result relies on the ability to chemically remove
with a well known efficiency a few atoms of \nuc{71}{Ge} produced by
neutrino interactions from $5 \times 10^{29}$ atoms of Ga.  To measure
this efficiency about 350~$\mu$g of stable Ge carrier is added to the
Ga at the beginning of each exposure, but even after this addition,
the separation factor of Ge from Ga is still 1 atom in 10$^{11}$.  We
have performed several auxiliary measurements which confirmed that the
technology of our experiment has the capability to extract
\nuc{71}{Ge} at this level.

An initial test was carried out in which Ge carrier doped with a known
number of \nuc{71}{Ge} atoms was added to a reactor holding 7~t of Ga.
Three successive extractions were made, and the number of \nuc{71}{Ge}
atoms in each extraction was determined by counting.  The results
\cite{ABD94} showed that the extraction efficiency for the natural Ge
carrier and the \nuc{71}{Ge} atoms did not differ.

A second experiment addressed the concern that the \nuc{71}{Ge} atom
from inverse beta decay may be created in an excited or ionized state
which results in the \nuc{71}{Ge} being tied up in a chemical form
from which we cannot efficiently extract.  A set of measurements
designed to test directly this question was carried out by observing
the beta decay of radioactive Ga isotopes in liquid Ga.  These
measurements \cite{ABD94} showed that the expected isotopes are formed
in the amounts anticipated at the 10\% level.

A test of all the experimental procedures including the chemical
extraction, counting, and the analysis technique was performed using
a 19.1~PBq (517~kCi) \nuc{51}{Cr} neutrino source.  The result,
expressed as the ratio of the measured \nuc{71}{Ge} production rate to
that expected due to the source strength, is $0.95 \pm 0.12$
\cite{ABD96,ABD98}.  This value provides strong verification that the
experimental efficiencies are as claimed and validates the fundamental
assumption in radiochemical experiments that the extraction efficiency
of atoms produced by neutrino interactions is the same as that of the
natural carrier.

\subsection{Tests of Analysis Hypotheses}

\begin{figure}
\begin{center}
\includegraphics[width=3.375in]{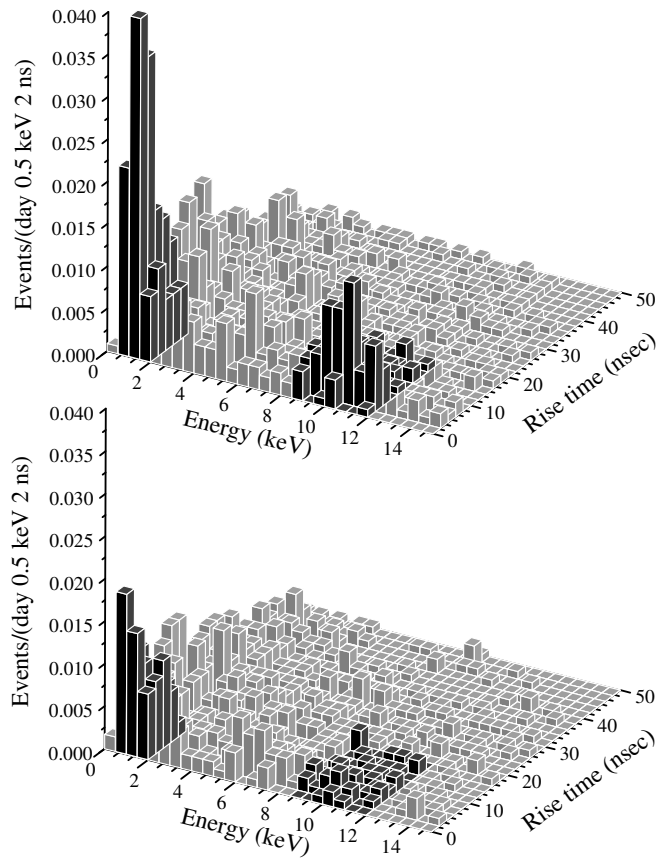}
\end{center}
\caption{Upper panel shows a histogram of energy vs rise time for all
events observed during the first 22.86~days after extraction for all
runs that could be counted in both $L$ and $K$ peaks (except May
1996).  The live time is 1169.9~days.  The approximate expected
location of the \nuc{71}{Ge} $L$ and $K$ peaks as predicted by
calibrations is shown darkened.  Lower panel shows the same histogram
for all events that occurred during an equal live time interval
beginning at day 100 after the time of extraction.}
\label{2d_hist}
\end{figure}

Direct visual evidence that we are really observing \nuc{71}{Ge} is in
Fig.\ \ref{2d_hist} which shows all events that survive the time cuts
and that do not have a NaI coincidence.  The expected location of the
\nuc{71}{Ge} $L$ and $K$ peaks is shown darkened.  These peaks are
apparent in the upper panel but missing in the lower panel because the
\nuc{71}{Ge} has decayed away.

\subsubsection{Time sequence}

\begin{figure}
\begin{center}
\includegraphics*[width=3.375in]{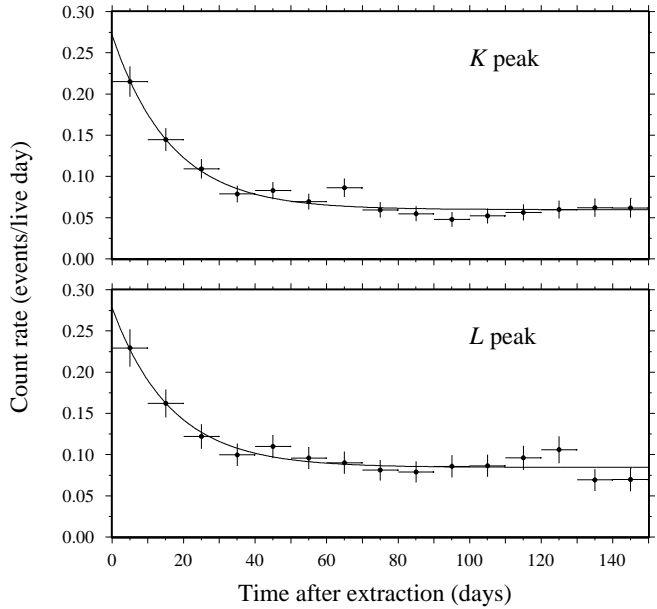}
\end{center}
\caption{Count rate for all runs from January 1990 in $L$ and $K$ peaks.  
The solid line is a fit to the data points with the 11.4-day half-life of
\nuc{71}{Ge} plus a constant background.  The vertical error bar on
each point is proportional to the square root of the number of counts
and is shown only to give the scale of the error.  The horizontal
error bar is $\pm$5~d, equal to the 10-day bin size.}
\label{countrate}
\end{figure}

A major analysis hypothesis is that the time sequence of observed
events for each run consists of the superposition of events from the
decay of a fixed number of \nuc{71}{Ge} atoms plus background events
which occur at a constant rate.  The quantity $Nw^2$ and the goodness
of fit probability inferred from it provide a quantitative measure of
how well the data fit this hypothesis.  These numbers are evaluated
for each data set and given in Table~\ref{recentruns}.  There are
occasional runs with low probability of occurrence, but no more of
these are observed than are expected due to normal statistical
variation.

This method can also be used to determine the goodness of fit for the
combined time sequence of all $L$ plus $K$ events from any combination
of runs.  For all SAGE runs this test yields $Nw^2 = 0.053$, with a
goodness of fit probability of $(72.0 \pm 4.5)\%$.  A visual
indication of the good quality of this fit is provided in
Fig.~\ref{countrate} which shows the count rate for all events in the
$L$ and $K$ peaks vs time after extraction.  An additional
quantitative indication that \nuc{71}{Ge} is being counted can be
obtained by allowing the decay constant during counting to be a free
variable in the maximum likelihood fit, along with the combined
\nuc{71}{Ge} production rate and all the background rates.  The best
fit half-life to all selected events in both $L$ and $K$ peaks is then
$9.7 ^{+1.5} _{-1.3}$~days, in agreement with the measured value
\cite{HAM85} of 11.43 days.

\subsubsection{Production rate sequence}

Another analysis hypothesis is that the rate of \nuc{71}{Ge}
production is constant in time.  By examination of
Fig.~\ref{ratevstime}, it is apparent that, within the large
statistical uncertainty for each run, there is no substantial
long-term deviation from constancy.

\begin{figure}[t]
\begin{center}
\includegraphics[width=3.375in]{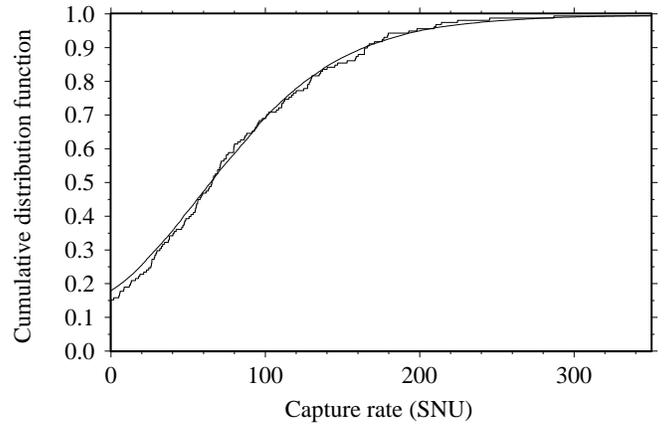}
\end{center}
\caption{Cumulative distribution function of the measured neutrino 
capture rate for all 158 SAGE data sets (jagged curve) and the 
expected distribution derived by 1000 Monte Carlo simulations of 
each set (smooth curve).  The capture rate in the simulations was 
assumed to be 70.8~SNU.}
\label{prod}
\end{figure}

Another way to consider this question is to use the cumulative
distribution function of the production rate $C(p)$, defined as the
fraction of data sets whose production rate is less than $p$.
Figure~\ref{prod} shows this distribution for all data sets and the
expected distribution from simulation, assuming a constant production
rate of 70.8~SNU.  The two curves parallel each other closely and can
be compared by calculating the $Nw^2$ test statistic \cite{CLE982}.
This gives $Nw^2$ = 0.337 whose probability is 11\%.

\begin{figure}
\begin{center}
\includegraphics[width=3.375in]{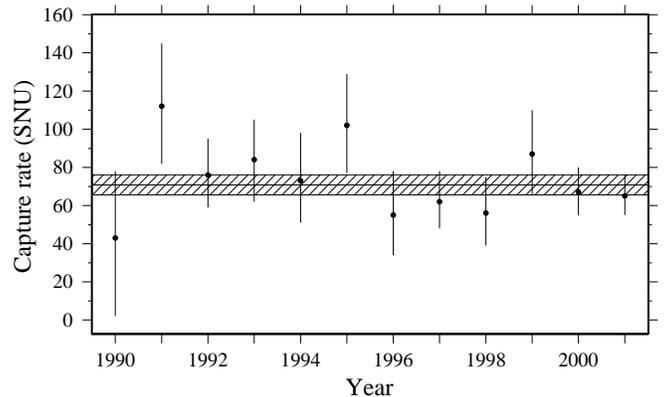}
\end{center}
\caption{Combined SAGE results for each year.  Shaded band is the 
combined best fit and its uncertainty for all years.  Error bars 
are statistical with 68\% confidence.}
\label{yearly}
\end{figure}

\subsection{Temporal Combinations of Data}

\begin{table*}[!htb]
\caption{Results of combined analysis of all runs during yearly,
monthly, and bimonthly intervals.  Runs are assigned to each time
period by their mean exposure time.  The accuracy of the probability
estimate is $\sim$4\%.}
\label{sbyyears}
\begin{tabular*}{\hsize}{l @{\extracolsep{\fill}} d d d d r
 @{\extracolsep{0pt}--} d @{\extracolsep{\fill}} c d}
\hline
\hline
          & \mco{Number } & \mco{Number of} & \mco{Number}  \\
Exposure  & \mco{of data} & \mco{candidate} & \mco{fit to}     & \mco{Best fit}
             & \multicolumn{2}{c}{68\% conf.}   &        & \mco{Probability} \\
interval  & \mco{sets   } & \mco{events   } & \mco{$^{71}$Ge}  & \mco{  (SNU) }
             & \multicolumn{2}{c}{range (SNU)}  & \mco{$Nw^2$} & \mco{   (\%) } \\
\hline
1990               &  5 &   43 &    4.9 &   43 &   2 &  78 & 0.260 &   9 \\
1991               &  6 &   59 &   25.5 &  112 &  82 & 145 & 0.120 &  17 \\
1992               & 13 &  145 &   39.8 &   76 &  59 &  95 & 0.047 &  68 \\
1993               & 15 &   97 &   33.2 &   84 &  62 & 105 & 0.199 &   6 \\
1994               & 10 &  155 &   24.1 &   73 &  51 &  98 & 0.027 &  95 \\
1995               & 13 &  210 &   37.7 &  102 &  77 & 129 & 0.041 &  82 \\
1996               & 10 &  120 &   19.3 &   55 &  34 &  78 & 0.067 &  51 \\
1997               & 16 &  183 &   35.7 &   62 &  48 &  78 & 0.057 &  62 \\
1998               & 12 &  189 &   26.7 &   56 &  39 &  75 & 0.064 &  60 \\
1999               & 14 &  114 &   40.8 &   87 &  66 & 110 & 0.068 &  33 \\
2000               & 22 &  235 &   62.2 &   67 &  55 &  80 & 0.102 &  29 \\
2001               & 22 &  173 &   64.4 &   65 &  55 &  76 & 0.050 &  70 \\
\hline
January            & 11 &  129 &   24.8 &   58 &  37 &  80 & 0.082 &  35 \\
February           & 12 &  101 &   25.5 &   60 &  44 &  77 & 0.045 &  74 \\
March              &  9 &  129 &   34.5 &  102 &  79 & 127 & 0.043 &  78 \\
April              &  9 &   80 &   16.9 &   54 &  37 &  73 & 0.072 &  39 \\
May                & 12 &  114 &   34.7 &   75 &  59 &  94 & 0.051 &  62 \\
June               & 11 &  101 &   33.6 &   79 &  58 & 102 & 0.175 &   5 \\
July               & 15 &  176 &   26.6 &   52 &  37 &  69 & 0.091 &  35 \\
August             & 15 &  161 &   38.7 &   78 &  60 &  96 & 0.058 &  51 \\
September          & 20 &  220 &   48.4 &   68 &  54 &  84 & 0.035 &  91 \\
October            & 17 &  169 &   40.3 &   73 &  57 &  91 & 0.080 &  45 \\
November           & 15 &  196 &   37.5 &   59 &  44 &  75 & 0.034 &  82 \\
December           & 12 &  147 &   46.4 &  105 &  84 & 127 & 0.040 &  89 \\
\hline
January+February   & 23 &  230 &   50.5 &   59 &  46 &  73 & 0.095 &  34 \\
March+April        & 18 &  209 &   49.2 &   75 &  61 &  91 & 0.026 & >99 \\
May+June           & 23 &  215 &   68.0 &   77 &  63 &  91 & 0.111 &  10 \\
July+August        & 30 &  337 &   65.4 &   65 &  53 &  78 & 0.075 &  50 \\
September+October  & 37 &  389 &   88.7 &   70 &  59 &  82 & 0.041 &  85 \\
November+December  & 27 &  343 &   84.3 &   78 &  66 &  91 & 0.042 &  85 \\
\hline
February+March     & 21 &  230 &   58.8 &   77 &  63 &  91 & 0.037 &  84 \\
April+May          & 21 &  194 &   50.8 &   66 &  54 &  79 & 0.049 &  60 \\
June+July          & 26 &  277 &   58.7 &   63 &  50 &  77 & 0.081 &  42 \\
August+September   & 35 &  381 &   87.1 &   72 &  61 &  84 & 0.043 &  84 \\
October+November   & 32 &  365 &   78.0 &   66 &  54 &  78 & 0.043 &  84 \\
December+January   & 23 &  276 &   73.6 &   84 &  70 &  99 & 0.059 &  65 \\
\hline
February+November  & 27 &  297 &   63.1 &   59 &  48 &  71 & 0.017 &  99 \\
March+October      & 26 &  298 &   75.1 &   84 &  71 &  99 & 0.062 &  66 \\
April+September    & 29 &  300 &   64.2 &   63 &  52 &  75 & 0.043 &  86 \\
May+August         & 27 &  275 &   73.3 &   77 &  64 &  89 & 0.045 &  75 \\
\hline
\hline
\end{tabular*}
\end{table*}

Neutrino oscillations can give a seasonal variation of the capture
rate for some values of the mass and mixing angle parameters
\cite{Berezinsky99,Fogli99}.  Other phenomena can also yield temporal
variations (see, e.g., \cite{Akhmedov99,Sturrock00}).  We thus give in
Table~\ref{sbyyears} the results of combining the SAGE runs in various
ways, monthly, bimonthly, and yearly.  There is no compelling evidence
for a temporal variation in any of these data divisions.  The yearly
results are plotted in Fig.~\ref{yearly} which shows that the rate has
been more or less constant during the data taking period.  Considering
only the statistical errors, a $\chi^2$ test against the hypothesis of
the constant rate of 70.8~SNU yields $\chi^2 = 6.6$, which, with 11
degrees of freedom, has a probability of 83\%.

Day-night effects can also produce large differences in the capture
rate between winter and summmer (see, e.g., \cite{Fogli99}).  Defining
summer as the $\pm1/4$-year interval centered on 21~June and winter as
the rest of the year, the winter minus summer difference in SAGE
capture rate is $R_{\text{W}} - R_{\text{S}} =
5.7^{+10.6}_{-10.3}$~SNU where the stated error is only statistical.
In our method of data analysis \cite{ABD99} we remove the known change
in rate caused by the Earth's orbital eccentricity, and thus in the
absence of neutrino oscillations the expected value for $R_{\text{W}}
- R_{\text{S}}$ is zero.  (If, rather than using the above
solstice-based definition, we define summer as the $\pm1/4$-year
interval centered on 5 July, the time of the aphelion, then
$R_{\text{W}} - R_{\text{S}} = 6.7^{+10.7}_{-10.3}$~SNU, also
consistent with a null rate.)

\section{The $\bm{\lowercase{pp}}$ Neutrino Flux}

One of the main purposes of the Ga experiment is to provide
information that leads to the experimental determination of the flux
of $pp$ neutrinos at the Earth.  In this Section we indicate the
present state of this measurement where we use only information from
the various solar neutrino experiments and assume that their reduced
capture rate compared to SSM predictions is due to neutrino
oscillations. \footnote{Note that the very restrictive luminosity
constraint \cite{Bahcall02} is not used here.}

By combining the results of SAGE, GALLEX, and GNO, the capture rate in
the Ga experiment is approximately $72 \pm 5$~SNU.  This rate is the
sum of the rates from all the components of the solar neutrino flux,
which we denote by [$pp$+$^7$Be+CNO+$pep+^8$B$|$Ga,exp], where ``exp''
indicates that this is a measured rate.  (We ignore the $hep$
contribution, as it is a negligible 0.05\% of the total in the SSM
calculation \cite{BAH00}.)

The only one of these flux components that is known is the \nuc{8}{B}
flux, measured by SNO to be [$^8$B$|$SNO,exp] = $(1.75 \pm 0.15)
\times 10^6$ electron neutrinos/(cm$^2$-s) \cite{MAC00}.  Since the
shape of the \nuc{8}{B} spectrum in SNO and SuperKamiokande is very
close to that of the SSM above 5~MeV and the cross section for Ga
rises steeply with energy, we can use the SNO flux and the cross
section for \nuc{8}{B} neutrinos with SSM shape ($2.40^{+0.77}_{-0.36}
\times 10^{-42}$ cm$^2$ [see \cite{BAH97,cross} for the cross sections
used here]) to conclude that the \nuc{8}{B} contribution to the Ga
experiment is [$^8$B$|$Ga,exp] = $4.2^{+1.4}_{-0.7}$ SNU.  Subtracting
this measured value from the total Ga rate gives
[$pp+^7$Be+CNO+$pep|$Ga,exp] = $67.8^{+5.1}_{-5.2}$ SNU.

The measured capture rate in the Cl experiment is
[$^7$Be+$^8$B+CNO+$pep|$Cl,exp] = $2.56 \pm 0.23$ SNU \cite{CLE98}.
(The $hep$ contribution will again be neglected as it is only 0.5\% of
the total in the SSM.)  Since the cross section in Cl is dominated by
neutrinos above 5~MeV, we can again use the SNO flux and the cross
section calculated for the SSM ($1.14^{+0.04}_{-0.04} \times 10^{-42}$
cm$^2$), and deduce that the contribution of \nuc{8}{B} to the Cl
experiment is [$^8$B$|$Cl,exp] = $2.00 \pm 0.18$ SNU.  Subtracting this
component from the total leaves [$^7$Be+CNO+$pep|$Cl,exp] = $0.56 \pm
0.29$ SNU, all of which is due to neutrinos of medium energy.

Neutrino oscillations have the effect of introducing an
energy-dependent survival factor to the fluxes predicted by the SSM.
For the medium-energy neutrinos this factor for the Cl experiment can
be approximated by the ratio of the measured rate to the SSM
prediction of [$^7$Be+CNO+$pep|$Cl,SSM] = $1.79 \pm 0.23$ SNU.  If we
assume that the survival factor varies slowly with energy, we find it
to be given by [$^7$Be+CNO+$pep|$Cl,exp]/[$^7$Be+CNO+$pep|$Cl,SSM] =
$0.31 \pm 0.17$.  Since the \nuc{7}{Be} contribution dominates, and it
is at a single energy, the error in this factor due to the assumption
that it is the same for all of these flux components can be estimated
by considering the contribution of the non-\nuc{7}{Be} components to
the total in the SSM, which is 36\%.  We thus increase the error from
0.17 to $0.17 + 0.31\times0.36 = 0.28.$

The relative contributions to the capture rate of the medium-energy
neutrinos are about the same in Ga as in Cl (75\% from \nuc{7}{Be} in
Ga compared to 64\% in Cl).  Thus it is reasonable to apply the
survival factor determined for Cl to the Ga experiment, i.e.,
[$^7$Be+CNO+$pep|$Ga,exp] = $(0.31 \pm 0.28) \times$
[$^7$Be+CNO+$pep|$Ga,SSM] = $14.4 \pm 13.0$ SNU.  We subtract this
contribution from the rate above and get the result for the measured
$pp$ rate in the Ga experiment [$pp|$Ga,exp] =
[$pp+^7$Be+CNO+$pep|$Ga,exp] - [$^7$Be+CNO+$pep|$Ga,exp] = $53 \pm 14$
SNU.

Since the cross section does not change appreciably over the narrow
range of Ga response to the $pp$ neutrinos, (0.23--0.42)~MeV, we
divide the capture rate by the SSM cross section for electron
neutrinos of $11.7^{+0.3}_{-0.3} \times 10^{-46}$ cm$^2$ and obtain
the measured electron neutrino $pp$ flux at Earth of $(4.6 \pm 1.2)
\times 10^{10}$/(cm$^2$-s).  Alternatively, if we divide the capture
rate by the cross section multiplied by the survival probability for
$pp$ neutrinos, which is 60\% for the favored LMA solution
\cite{BAH011} assuming no transitions to sterile neutrino flavors, we
receive the rate of $pp$ neutrino emission in the solar fusion
reaction of $(7.6 \pm 2.0) \times 10^{10}$/(cm$^2$-s), in agreement
with the SSM calculation of $(5.95 \pm 0.06)\times 10^{10}$/(cm$^2$-s)
\cite{BAH00}.  The major component of the error in the $pp$ flux
measurement is due to the poor knowledge of the energy-dependent
survival factor.

Several approximations were made in arriving at this value, whose
nature cannot be easily quantified, so perhaps the error given here is
somewhat underestimated.  Nonetheless, as mentioned in the
Introduction, it will be possible to reduce the error in this flux
greatly when the region of mass and mixing angle parameters is better
determined, as should be done by the KamLAND experiment, and when the
\nuc{7}{Be} flux is directly measured, as anticipated by
Borexino~\cite{Bil01}.  The dominant error should eventually be due to
the inaccuracy of the Ga measurement itself, and hence we are seeking
to reduce our statistical and systematic errors.

\section{Summary and Conclusions}

The methods and analysis of the SAGE experiment have been summarized
and results for 92 extractions during 12 years of operation from
January 1990 through December 2001 have been presented.  The measured
capture rate is $70.8^{+5.3}_{-5.2}$~SNU where the uncertainty is
statistical only.  Analysis of all known systematic effects indicates
that the total systematic uncertainty is $^{+3.7}_{-3.2}$~SNU, less
than the statistical error.  Finally we have examined the counting
data and shown that there is good evidence that \nuc{71}{Ge} is being
counted, that the counting data fit the analysis hypotheses, and that
the counting data are self-consistent.

The SAGE result of 70.8~SNU represents 55\% of SSM
predictions~\cite{BAH00}.  Given the extensive systematic checks and
auxiliary measurements that have been performed, especially the
\nuc{51}{Cr} neutrino source experiment \cite{ABD96,ABD98}, this
6.0$\sigma$ reduction of the solar neutrino capture rate compared to
SSM predictions is very strong evidence that the solar neutrino
spectrum below 2~MeV is significantly depleted, as has been proven for
the \nuc{8}{B} flux by the Cl, Kamiokande, and SNO experiments.
The SAGE result is even somewhat below the astrophysical minimum
capture rate of $79.5^{+2.3}_{-2.0}$~SNU \cite{BAH97}.

Several recent phenomenology papers (see, e.g.,
\cite{BAH012,KRA01,GON01}) discuss the combined fit of all solar
neutrino experiments.  Their conclusion is that the electron neutrino
oscillates into other species and the best fit is to the LMA region of
Mikheyev-Smirnov-Wolfenstein (MSW) oscillations.  To more precisely
determine the oscillation parameters in the solar sector will require
additional data, especially from experiments sensitive to the
low-energy neutrinos.  In this vein, SAGE continues to perform regular
solar neutrino extractions every four weeks with $\sim$50~t of Ga and
will continue to reduce its statistical and systematic uncertainties.

\section*{Acknowledgments}

We thank J.~N.~Bahcall, M.~Baldo-Ceolin, G.~T.~Garvey, W.~Haxton,
V.~A.~Kuzmin, V.~V.~Kuzminov, V.~A.~Matveev, L.~B.~Okun,
R.~G.~H.~Robertson, V.~A.~Rubakov, A.~Yu.~Smirnov, A.~N.~Tavkhelidze,
and many members of GALLEX and GNO for their continued interest and
for stimulating discussions.  We greatly appreciate the work of our
prior collaborators O.~L.~Anosov, O.~V.~Bychuk, M.~L.~Cherry,
R.~Davis, Jr., I.~I.~Knyshenko, V.~N.~Kornoukhov, R.~T.~Kouzes,
K.~Lande, A.~V.~Ostrinsky, D.~L.~Wark, P.~W.~Wildenhain, and
Yu.~I.~Zakharov.  We acknowledge the support of the Russian Academy of
Sciences, the Institute for Nuclear Research of the Russian Academy of
Sciences, the Russian Ministry of Science and Technology, the Russian
Foundation of Fundamental Research under grant No.\ 96-02-18399, the
Division of Nuclear Physics of the U.S. Department of Energy under
grant No.\ DEFG03-97ER41020, and the U.S. Civilian Research and
Development Foundation under award No.\ RP2-159.  This research was
made possible in part by grant No.\ M7F000 from the International
Science Foundation and grant No.\ M7F300 from the International
Science Foundation and the Russian Government.

\end{document}